# Privacy-Preserving Identity and Access Management in Multiple Cloud Environments: Models, Issues, and Solutions

Alfredo Cuzzocrea*
*iDEA Lab*
*University of Calabria*
Rende, Italy
& *Dept. of Computer Science*
*University of Paris City*
Paris, France
alfredo.cuzzocrea@unical.it

Islam Belmerabet
*iDEA Lab*
*University of Calabria*
Rende, Italy
ibelmerabet.idealab.unical@gmail.com

*Abstract*—This paper focuses the attention on *privacy-preserving identity and access management in multiple Cloud environments*, which is an annoying problem in the modern *big data era*. Within this conceptual context, the paper describes contemporaneous models and issues, and put the basis for future solid solutions. Finally, we provide a summary table where we embed an innovative taxonomy of state-of-the-art research proposals in the reference scientific field.

## I. INTRODUCTION

Nowadays, *Cloud computing* offers dynamic and on-demand access to computing resources (e.g., [1]), making it a key technology across industries. However, the increasing reliance on Cloud services has heightened concerns around security, particularly in *Identity and Access Management* (IAM). IAM systems, largely controlled by *Cloud Service Providers* (CSPs), are essential for ensuring data protection, yet their centralized nature often falls short in addressing *user-specific needs* for *fine-grained control* (e.g., [2]). This issue becomes more prominent as Cloud environments expand into *private*, *public*, and *hybrid/federated Clouds*, each with unique security challenges.

*Cloud service models*, i.e., *Software-as-a-Service* (SaaS), *Platform-as-a-Service* (PaaS) and *Infrastructure-as-a-Service* (IaaS), allow for flexible deployment but necessitate robust authentication and authorization systems due to risks like identity theft, data leakage, and loss of control over data stored by third-party providers. Moreover, Cloud computing is built on a *service-oriented architecture* that may provide *Database-as-a-Service* (DbaaS), *Identity-as-a-Service* (IDaaS), and *Anything-as-a-Service* (XaaS) (e.g., [3,4]).

In the context of Cloud computing, ineffective mechanisms lead to several security problems, including *risk management*, *compliance*, *data security*, *privacy*, *transparency,* and *data leakage* (e.g., [5-8]). The intricacy of Cloud systems and the security issues. For instance, issues with transparency and loss of control arise when user data are processed and stored by CSPs or when handled outside of corporate walls. Despite these advances, security issues persist, particularly around data transparency, compliance, and user trust. Companies are hesitant to share sensitive identity data due to these vulnerabilities, delaying Cloud adoption. To address these concerns, IAM systems offer authentication, authorization, and access management services, ensuring only authorized users can access Cloud resources [9].

Indeed, *digital identity management* plays a crucial role in mitigating these risks. However, managing multiple digital identities and credentials presents challenges for users, necessitating streamlined identity management across trust domains. *Open identity management systems* (e.g., [10,11]) have emerged to address this, enabling the exchange of identity information across domains securely.

*Digital Identity Management* involves constructing and regulating a user digital identity to reflect *real-life identity* within *online systems*, by enabling activities like *online banking*, *shopping* and *social networking* (e.g., [12,13]). Digital identity management ensures that *Service Provider* (SP) can verify user identity for personalized services and accountability. Digital identities are typically subsets of a user real-life attributes but managing multiple identities and credentials can be burdensome for users. Open identity management approaches emerged to simplify this by enabling secure identity information sharing across domains through *Identity Providers* (IdPs), which store user data and issue "*Information Cards*" for *user-IdP interaction*. These Information Cards help users connect with their IdPs to access identity attributes and security tokens for verification.

Furthermore, Cloud environments bring new security challenges for digital identity management, especially when *third-party vendors* manage data. IAM systems address these concerns by ensuring secure authentication, authorization and access control for *Cloud services* [14]. Through IAM, only authorized individuals with specific access rights can handle sensitive *Cloud-stored data*, which offers better protection against *security risks* [15]. Many businesses leverage IAM systems to enhance security for Cloud-stored data, by ensuring user identities and access privileges are managed effectively within Cloud infrastructures [16].

Following these considerations, this paper addresses the need for a structured approach for *privacy-preserving digital identity management in multi-Cloud environments*, a topic that remains *under-explored* in active literature. It aims to fill this gap by presenting a comprehensive survey of models, methods and techniques designed to secure digital identities across diverse Cloud platforms. Organized around seven core topics, *Authentication and Authorization, Identity Verification and Risk Management, Attribute Management, Security and*

*Privacy*, *Audit and Compliance*, *User Lifecycle Management*, and *Interoperability and Federation*, this study provides a detailed examination of these aspects particularly to achieve robust security, privacy, and interoperability across multi-Cloud platforms.

The remaining part of this paper is organized as follows: in Section II, we provide a formal background on the privacy-preserving digital identity management discipline over multi-Cloud environments. Section III introduces the seven specialized topics that form the foundation of our study. Finally, in Section IV, we present our conclusions and outline potential areas for future work.

## II. IDENTITY MANAGEMENT OVER MULTIPLE CLOUD: CONCEPTUAL CONTEXT

In this Section, we provide a comprehensive and contextual background on digital identity management in the context of multi-Cloud.

In order to guarantee consistent security and compliance, digital identity management in multi-Cloud refers to the safe management and control of user identities and access across various *Cloud platforms*. By retaining user identities across different Cloud services, protecting user privacy and reducing the exchange of sensitive data between Clouds, multi-Cloud privacy-preserving digital identities are accomplished.

Identity is an *entity information*. Digital identity is what we obtain when we use a digital format to express the physical identity [17]. "*Digital identity*" shall be referred to as "*identity*" from now on. Identity is often connected to a person, but it may also refer to programs, services, gadgets and systems that represent a person. It may also be connected to non-human beings like a beloved dog and inanimate things like a building, a vehicle transporting food, or a nuclear warhead. This research scope is restricted to the identity of human subjects. Additionally, the diversity and dynamic nature of identification information presented by people present a unique and fascinating set of identity difficulties. Information about an individual identification may be utilized for several objectives. One can utilize their mother maiden name as an authenticator, their Visa card number as an identifier and their *Personal Identification Number* (PIN) as an authorizer while withdrawing cash. Since identities are the foundation of every organization, they must be safeguarded against abuse [18,19].

Digital identity formation, usage and termination are all governed by procedures and rules that are part of identity management. It makes data, systems and applications accessible in an *effective*, *secure*, and *safe* manner [20]. Identity management is critical to system security, building user trust, managing users, keeping an eye on illicit and criminal activity, ensuring legal compliance, supporting auditing and reporting procedures, provisioning, access control, resource protection, confidentiality and information protection. Identity management techniques control an identity temporal and spatial constraints. For instance, a student receives resources and rights depending on their identification when she joins a research team; they are either regained or revoked when they depart the team.

Most identities on the Internet are centralized. There is only one organization that owns and maintains the user credentials. However, there are shortcomings with this separate identity repository strategy. *Identity providers* have the ability to cancel or abuse users data; they do not possess them. *Federated identity management solutions*, on the other hand, have the ability to offer permission and authentication across systems [21,22]. Contractual agreements on data ownership are necessary, as are agreements that an identity at one provider be recognized by other providers. There is no need for business directory integration because user accounts are controlled independently by identity providers. Because credentials are propagated on demand rather than duplicated, the security risk is reduced. The implementation of this strategy is comparatively more complicated and calls for appropriate trust and agreement across internet providers. Lastly, the idea of "*self-sovereign identity*" holds that people need to have authority over their own digital identities. Without depending on a central repository of identity data, individuals and organizations can keep their identity data on their own devices and communicate their identity to others who need to confirm it. Because it is not connected to any specific silo, the user has complete control over the security and mobility of their data. Self-sovereign identity, according to the *Sovrin Foundation* [23], is like an Internet for identities where anybody may use it, enhance it and no one owns it.

The capacity to manage how information about oneself is handled is known as *privacy* [24,25]. Our constitution has always placed a high priority on privacy, with many revisions protecting people [26]. With the introduction of commercial computers in the 1960s, concerns over electronic records became a hot subject for congressional discussions. The US government has passed several pieces of legislation since then. Establishing guidelines and putting procedures in place to abide by privacy laws constitute privacy management. The *Privacy Act of 1974* [24] forbids any government system from disclosing electronic records on an individual without the person to whom the records relate having given written approval. This includes records about pupils in educational settings. There are several other regulations for various fields, including finance, auditing, homeland security and so forth.

*Federated Cloud* and *multi-Cloud* are two delivery methods for multiple Clouds. When many providers come to an agreement to work together, the user is not informed that resources are being accessed from another Cloud in federated Clouds. However, because there is no prior agreement between participating Clouds and cooperation is developed at runtime in accordance with needs, multi-Clouds offer a means for dynamic collaboration between different Clouds. Additionally, in multi-Cloud environments, users are directly in charge of the provisioning of services from different Clouds and possess knowledge of all linked Clouds, which can prove advantageous for both users and organizations. The typical Cloud landscape is altered when users access services from numerous Cloud providers in a multi-Cloud context. It is anticipated that multi-Clouds will become the standard in business for managing large volumes of data since they provide more flexibility, creativity and intensive cooperation [26].

*Blockchain* is a *peer-to-peer* (P2P) distributed database, also known as a ledger, that keeps track of a list of ever-expanding records, or blocks, that are connected and secured, often through the use of public key cryptography [27]. Using *blockchain* technology, instead of adding to the centralized database as in a typical centralized system, new information is added to a block and made available to all nodes in a distributed network. A hash value created, often using the safe

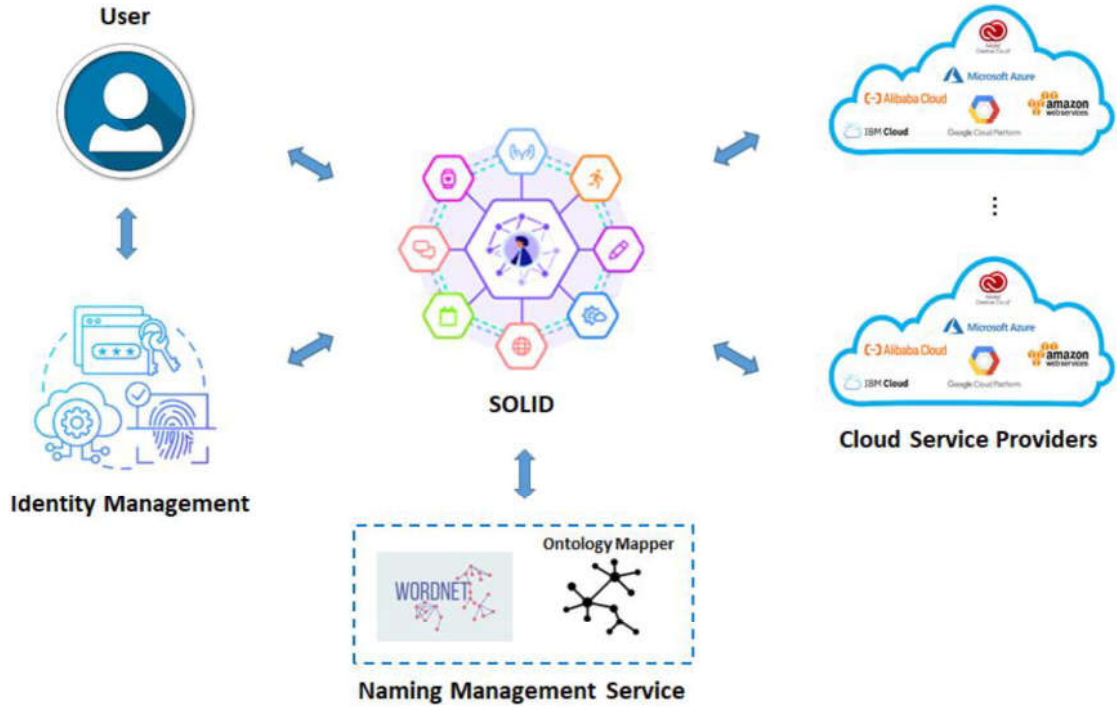

Fig. 1. Case Study: Supporting Digital Identity Management over Multi-Cloud Settings

hash cryptographic technique 256 bits (SHA256), identifies each block in a *blockchain* [28].

The following block (child) is connected to and stores the hash value of the current block header (parent) [29]. As a result, if the content of any block is altered, its hash will likewise be modified and this modification will spread throughout the network to render that block invalid [28]. This technique explains why blockchain technology is decentralized and disseminated without the need for a *middleman* or other reliable third party. To digitally sign and validate transactions, blockchain participants are granted private keys. Solid [30] is a standard that permits the safe storage of user data in what are known as *Personal Online Datastores* (Pods).

Pods are data-secured versions of private web servers. A user may manage which entities, apps and Cloud services can access data that is stored in their Pod. Standard, open and compatible data formats and protocols are used for the storage and retrieval of all data within a Solid Pod. Solid makes use of a standard, shared vocabulary that allows many applications to comprehend objects and their relationships with one another. Solid now has the special capability of enabling many applications to use the same data [31].

Fig. 1 provides a reference architecture for the case study related to the actual multi-Cloud setting. As shown in Fig. 1, several layers can be identified in the reference architecture. In the following, we highlight these layers:

- *User Application Layer*;
- *Solid Framework Layer*;
- *Cloud Service Provider Layer*.

III. IDENTITY MANAGEMENT OVER MULTIPLE CLOUD: THE RELEVANT PILLARS

In this Section, we introduce the main topics, under which we select several reference research works to describe and detail. The specialized topics are introduced to help frame the scope of this work.

*A. Authentication and Authorization*

*Authentication* and *authorization* are fundamental components of digital identity management, ensuring secure access to digital systems. Authentication is the first step, involving the verification of a user identity through various methods such as *passwords*, *biometrics* and *cryptographic keys*. This process establishes trust before allowing users access to a system. Authorization follows authentication, by defining what actions and resources the authenticated user can access. It enforces specific access policies based on attributes like *user roles* and *group memberships*, which ensures that users can only interact with data or functionalities aligned with their permissions. Together, these processes protect against unauthorized actions, leading to preserving *data privacy* and supporting security policies.

The relationship between authentication and authorization is deeply interconnected, as proper authorization relies on successful authentication. It determines what activities an entity, such as a user or device, is permitted to perform, including reading or writing data or controlling actuators. In Cloud computing, *Single Sign-On* (SSO) simplifies authentication by allowing users to access multiple systems within a single session [32,33]. Strong authentication methods, such as two-factor authentication, enhance security, while robust authorization models like *Role-Based Access Control* (RBAC) and *Attribute-Based Access Control* (ABAC) ensure that only authorized users access critical resources [34]. A lack of effective authentication or authorization mechanisms can lead to serious security risks, including *privilege escalation*, *data altering* or *service hijacking*.

*B. Identity Verification and Risk Management*

*Identity Verification and Risk Management* are key components of digital identity management, by ensuring security, trust and dependability in online interactions and transactions. Identity verification confirms the legitimacy of a user claimed identity through methods like personal information validation, biometric authentication and multi-factor authentication. These techniques help reduce fraud and unlawful access, especially in sensitive sectors like banking, healthcare and e-commerce, where compliance and customer trust are critical. Risk management identifies and mitigates threats by monitoring user behaviors and transaction patterns in a real-time manner and applies security measures based on *user risk profiles*. High-risk users face stricter controls, while low-risk users enjoy smoother access. Effective identity management mitigates operational risks such as insider threats or poor access control by managing and auditing access rights. Together, these processes form a robust foundation for digital identity management, by ensuring a *tradeoff* between user security and convenience to establish trust and compliance.

In addition to securing online interactions, Identity Verification and Risk Management are vital for regulatory compliance and enhancing customer trust, especially in industries governed by strict data protection laws. Effective identity management not only reduces the risk of cyber threats but also helps organizations avoid fines and reputational damage caused by data breaches. Advances in artificial intelligence and machine learning are transforming risk management, by providing dynamic and context-aware systems that adapt to emerging threats by identifying unusual patterns and predicting risks. Implementing robust authentication mechanisms, like biometrics and behavioral analysis, further reduces reliance on outdated password-based systems, which are prone to vulnerabilities. As digital ecosystems continue to expand, the role of identity verification and risk management becomes even more crucial, supporting the scalability of secure and seamless user experiences across various platforms and services.

*C. Attribute Management*

*Attribute Management* is a critical aspect of digital identity management, responsible for handling and controlling information linked to *individual identity*, such as name, age and access rights. This process involves collecting, validating, storing and delivering these attributes to ensure secure access to resources and services. The data is safely stored in an *Identity Store* and can be exchanged with service providers (SP) through established protocols like *SAML* and *OpenID Connect*. This ensures that only the necessary information is shared, following the principle of "*minimum disclosure*" to enhance privacy and security (e.g., instead of sharing a full birthdate, it may only reveal that the individual is above 18 years old). In order to avoid misunderstandings, standardized procedures for attribute exchange must ensure that both the attribute provider and recipient understand the structure and format of the attributes.

The assurance of attributes also plays a key role, by requiring evaluation of both the quality of the data and the link between the digital identity and the attribute. International initiatives, like *STORK2.0 project* [35] and *European Thematic Network SSEDIC* [36], have developed standards for determining the assurance levels of attributes, though further global standardization is needed. Effective attribute management is crucial for systems like SSO, streamlining user access across services by centralizing management. This not only improves user experience but also simplifies the administrative process of identity management, ensuring secure, *privacy-compliant access* to resources.

*D. Security and Privacy*

In digital identity management, *security* and *privacy* are two crucial and interconnected principles that protect user data and build trust in online interactions. Security involves a comprehensive set of measures, such as authentication, encryption, access controls and threat detection, designed to safeguard user identities and personal information from unauthorized access and cyber threats. Effective security strategies aim to prevent data breaches and identity theft, ensuring that only legitimate users can access authorized data and services. Meanwhile, privacy focuses on protecting individual personal data, requiring user consent for data collection and usage in accordance with legal and ethical standards. Key privacy measures include data anonymization, user-controlled access and transparency in data handling, all of which help maintain user rights and foster trust.

The interplay between *security* and *privacy* forms a robust foundation for digital identity management, enabling individuals to engage confidently in online activities while controlling their personal information [37]. Varying definitions and societal perspectives, highlighting the importance of protecting against identity theft and related risks, such as unauthorized access and fraud, underscore the significance of privacy. Addressing privacy intricacies necessitates an interdisciplinary approach that considers legal frameworks, societal norms and technological factors [38]. Integrating privacy from the outset, often termed "*privacy by design*", which ensures that privacy requirements are embedded within digital identity management systems from the beginning, thereby enhancing privacy-preserving and ensuring that these considerations are integral to system design and functionality [39].

*E. Audit and Compliance*

*Audit and compliance* are essential components of digital identity management, ensuring accountability, transparency and adherence to security standards and regulations. Auditing involves monitoring, recording and analyzing identity-related activities, such as authentication attempts and access requests, to track compliance with regulatory frameworks. Audit logs provide historical records that help organizations assess security measures, investigate suspicious behavior and support incident response. Compliance, on the other hand, focuses on adhering to laws and standards like GDPR, HIPAA and PCI-DSS that require organizations to implement security controls, conduct risk assessments and regularly audit their identity management systems to ensure alignment with these regulations. Non-compliance can lead to legal penalties and damage to organizational reputation.

The relationship between audit and compliance is symbiotic, as audits allow organizations to demonstrate compliance, identify improvement areas and address security vulnerabilities proactively. For *Cloud-based identity management systems* [40], auditing and compliance features are critical for establishing trust and reducing risks such as insider threats and *data loss*. Systems that meet international security standards are considered highly reliable. In sectors like banking and healthcare, regulations mandate stringent controls and improving identity management practices, including provisioning, authentication and access control, can reduce costs, enhance efficiency and create new opportunities for secure access to information and services.

TABLE I. PROPERTY TO PILLAR MAPPING

| Property \ Pillar | Authentication and Authorization | Identity Verification and Risk Management | Attribute Management | Security and Privacy | Audit and Compliance | User Lifecycle Management | Interoperability and Federation |
|---|---|---|---|---|---|---|---|
| Authentication | ✓ | ✓ | ✓ | ✓ | | ✓ | |
| Access Control | ✓ | ✓ | | ✓ | | ✓ | |
| Privacy Preserving | ✓ | | ✓ | ✓ | ✓ | | |
| Encryption | ✓ | | ✓ | ✓ | | | |
| Identity Management | ✓ | ✓ | ✓ | | ✓ | ✓ | ✓ |
| Authorization | ✓ | | | ✓ | | ✓ | |
| Identity Provider | ✓ | ✓ | ✓ | | ✓ | ✓ | ✓ |
| Trust Management | ✓ | ✓ | ✓ | ✓ | ✓ | | |
| Interoperability | | | | | | | ✓ |
| Federated Identity Management | ✓ | ✓ | ✓ | | ✓ | ✓ | ✓ |

*F. User Lifecycle Management*

*User Lifecycle Management* in digital identity management involves managing a user identity and access rights throughout their entire relationship with an organization, from *onboarding* to *offboarding*. The process includes four key phases: (*i*) *Provisioning*, where a user digital identity is created and access rights are assigned based on their role, often using automated tools; (*ii*) *Maintenance*, which involves ongoing updates to user attributes, roles and privileges as needed; (*iii*) *Access Review and Recertification*, where regular reviews assess whether access rights are still necessary to ensure alignment with least privilege principle; (*iv*) *De-provisioning*, where access rights are revoked when a user leaves the organization or no longer requires specific access, by preventing unauthorized access and limiting security risks.

User lifecycle management is crucial for maintaining security, data privacy and compliance with regulations, reducing risks such as orphaned accounts and access creep. A robust *Identity Management System* supports user lifecycle management by automating identity-related tasks, integrating tools for authentication, provisioning and access control and managing user roles and credentials. This system enhances operational efficiency and security, ensuring that resource access is properly managed based on job functions [41].

*G. Interoperability and Federation*

*Interoperability* and *federation* are pivotal concepts in digital identity management, enabling seamless and secure access across various systems, services and organizations. Interoperability refers to the ability of different identity management systems, applications, or components to work together effectively, even if they are developed by different vendors or operate on diverse platforms. This allows for the integration and sharing of identity data, authentication methods and access control policies among different systems and services. Users can access multiple resources using a single set of credentials or attributes, enhancing their experience and simplifying administrative processes, especially in complex IT environments.

Federation, on the other hand, enables the secure sharing of identity and access information between organizations, domains, or applications. It permits users to access resources in one domain using credentials from another, eliminating the need for separate accounts. Federation relies on standardized protocols and trust agreements, typically involving an identity provider and a service provider. The IdP authenticates the user and shares the necessary identity attributes with the SP, facilitating single sign-on capabilities across different domains or services.

Combined, interoperability and federation enhance user convenience, security and data privacy, making them essential components of modern identity and access management ecosystems. Organizations require identity and access management solutions that use standard interfaces to establish seamless identities across various technologies from different vendors, reducing integration costs. Some of the principal relevant standards include *Lightweight Directory Access Protocol* (LDAP) [42], *Service Provisioning Mark-up Language* (SPML) [43], SAML [44], and *Extensible Access Control Mark-up Language* (XACML) [45]. *Cross-domain federation* allows applications or products from different vendors to share information about authenticated users without requiring re-authentication for each service, which facilitates authentication and authorization processes.

Tab. I shows an analysis of the properties associated with each pillar of digital identity management over multi-Cloud. Overall, the Table defines a taxonomy of state-of-the-art research proposals in the reference scientific field, which is one of the main contributions of our research.

## IV. Conclusions and Future Work

In this paper, we provide a comprehensive and critical survey of privacy-preserving digital identity management models, methods, and techniques within multi-Cloud environments. By focusing on seven key areas, *Authentication and Authorization*, *Identity Verification and Risk Management*, *Attribute Management*, *Security and Privacy*, *Audit and Compliance*, *User Lifecycle Management*, and *Interoperability and Federation*, we structured our analysis to address the current state of research and identify future challenges within each of these pillars.

Future work aims at extending the proposal beyond privacy-preserving digital identities in multi-Cloud. As to explore other dimensions of digital identity management, all with an emphasis on maintaining privacy. As well as addressing and dealing with research challenges driven by emerging *big data trends* (e.g., [46-52]).

## Acknowledgment

This work was partially supported by project SERICS (PE00000014) under the MUR National Recovery and Resilience Plan funded by the European Union - NextGenerationEU.